%% file: neurips_2025.tex
\documentclass{article}

    \PassOptionsToPackage{numbers, compress}{natbib}

\usepackage[preprint]{neurips_2025}

\usepackage[utf8]{inputenc} 
\usepackage[T1]{fontenc}    
\usepackage[hidelinks]{hyperref}       
\usepackage{url}            
\usepackage{booktabs}       
\usepackage{amsfonts}       
\usepackage{nicefrac}       
\usepackage{microtype}      
\usepackage[table,xcdraw]{xcolor}
\usepackage{amsmath,amssymb}
\usepackage{array}
\usepackage{ragged2e}

\usepackage{tabularx}
\usepackage{textcomp}
\usepackage[T1]{fontenc}
\usepackage{lmodern}
\usepackage{enumerate}
\usepackage[utf8]{inputenc}
\usepackage{parskip}
\usepackage{enumitem}
\usepackage{xcolor}
\usepackage{geometry}
\usepackage[nospace]{varioref}
\usepackage{amsmath}

\usepackage{graphicx}
\usepackage{caption}
\usepackage{subcaption}

\usepackage{colortbl}
\usepackage{multirow}

\usepackage{booktabs}    
\usepackage{siunitx}     
\usepackage{adjustbox}   
\newcolumntype{L}[1]{>{\raggedright\arraybackslash}p{#1}}  
\newcolumntype{C}[1]{>{\centering\arraybackslash}p{#1}}

\definecolor{legalblue}{RGB}{33, 100, 200}
\definecolor{technicalorange}{RGB}{230, 110, 0}
\definecolor{humangreen}{RGB}{0, 150, 80}
\definecolor{orgpurple}{RGB}{128, 0, 128}
\definecolor{lightgray}{gray}{0.95}

\definecolor{DarkRed}{HTML}{922B21}
\definecolor{DarkYellow}{HTML}{d4ac0d}
\definecolor{DarkGreen}{HTML}{229954}

\newcommand{\taglegal}{\textcolor{legalblue}{\emph{\textbf{(Legal)}}}}
\newcommand{\tagtech}{\textcolor{technicalorange}{\emph{\textbf{(Technical)}}}}
\newcommand{\taghuman}{\textcolor{humangreen}{\emph{\textbf{(Human-Centered)}}}}
\newcommand{\tagorg}{\textcolor{orgpurple}{\emph{\textbf{(Organizational)}}}}

\newcommand{\taglegalnop}{\textcolor{legalblue}{\emph{\textbf{Legal:}}}}
\newcommand{\tagtechnop}{\textcolor{technicalorange}{\emph{\textbf{Technical:}}}}
\newcommand{\taghumannop}{\textcolor{humangreen}{\emph{\textbf{Human-Centered:}}}}
\newcommand{\tagorgnop}{\textcolor{orgpurple}{\emph{\textbf{Organizational:}}}}

\newcommand{\taglegaldef}{\textcolor{legalblue}{\emph{\textbf{Legal }}}}
\newcommand{\tagtechdef}{\textcolor{technicalorange}{\emph{\textbf{Technical }}}}
\newcommand{\taghumandef}{\textcolor{humangreen}{\emph{\textbf{Human-Centered }}}}
\newcommand{\tagorgdef}{\textcolor{orgpurple}{\emph{\textbf{Organizational }}}}

\title{Explainable AI Systems Must Be Contestable: Here's How to Make It Happen}

\author{%
  Catarina Moreira \\
  Data Science Institute\\
  UTS Australia\\
  \texttt{catarina.pintomoreira@uts.edu.au} 
  \And
  Anna Palatkina \\
  IT:U Austria \\
  \texttt{anna.palatkina@it-u.at} 
  \And
  Dacia Braca \\
  IT:U Austria\\
  \texttt{dacia.braca@it-u.at} \\
  \And
  Dylan M. Walsh \\
  Data Science Institute\\
  UTS Australia\\
  \texttt{dylan.m.walsh@student.uts.edu.au} 
  \And
  Peter J. Leihn \\
  Data Science Institute\\
  UTS, Australia\\
  \texttt{peter.leihn@uts.edu.au}
  \And
  Fang Chen \\
  Data Science Institute\\
  UTS,  Australia\\
  \texttt{fang.chen@suts.edu.au} 
  \And
  Nina C. Hubig \\
  IT:U Austria\\
  \texttt{nina.hubig@it-u.at} \\
}

\begin{document}

\maketitle

\begin{abstract}


As AI regulations around the world intensify their focus on system safety, contestability has become a mandatory, yet ill-defined, safeguard. In XAI, "contestability" remains an empty promise: no formal definition exists, no algorithm guarantees it, and practitioners lack concrete guidance to satisfy regulatory requirements. Grounded in a systematic literature review, this paper presents the first rigorous formal definition of contestability in explainable AI, directly aligned with stakeholder requirements and regulatory mandates. We introduce a modular framework of by-design and post-hoc mechanisms spanning human-centered interfaces, technical architectures, legal processes, and organizational workflows. To operationalize our framework, we propose the Contestability Assessment Scale, a composite metric built on more than twenty quantitative criteria. Through multiple case studies across diverse application domains, we reveal where state-of-the-art systems fall short and show how our framework drives targeted improvements. By converting contestability from regulatory theory into a practical framework, our work equips practitioners with the tools to embed genuine recourse and accountability into AI systems.


\end{abstract}

\section{Introduction}

Artificial intelligence (AI) systems are increasingly entrusted with high-stakes decisions in domains such as credit scoring, medical diagnosis, and autonomous vehicles \cite{sreedharan2020bridging, srivastava2024ai, marengo2023challenges, alfrink2023contestablecars}.  
In recognition of the potential for errors, biases, and opaque decision-making, regulators worldwide (including the European Union’s AI Act (Regulation (EU) 2024/1689) \cite{meltzer2022european} and Australia’s voluntary AI ethics framework \cite{lyons2021conceptualising}), have begun to mandate \emph{contestability}, i.e., mechanisms that enable stakeholders to dispute, review, and revise automated outcomes \cite{walmsley2021artificial}. 

Despite growing regulatory emphasis on contestability, the explainable AI (XAI) community still lacks a clear and operational understanding of what contestability entails \cite{alfrink2023contestable}. While multiple definitions exist across disciplines, there is no unified, versatile, and widely accepted conceptualization—let alone a translation into practical mechanisms. 

Existing work on interpretability and post hoc explanations does not translate into actionable dispute processes; there is no consensus on the dimensions of contestability, no quantitative metrics to assess it, and practitioners are left without design guidance to meet compliance or ethical obligations \cite{lyons2021conceptualising, geiger2024making, aloisi2024regulating}.

To the best of our knowledge, this paper is the first to confront this misalignment of contestability between theoretical AI regulations and real‐world XAI practice. Grounded in a systematic literature review and enriched by insights from human–machine interaction, we offer a rigorous, multidimensional definition of contestability tailored to explainable AI.  Building on this foundation, we present a modular framework of by‐design and post‐hoc mechanisms spanning human‐centered interfaces, technical architectures, legal processes, and organizational workflows.  To make contestability measurable, we introduce the \emph{Contestability Assessment Scale}, a composite metric based on over twenty quantitative criteria.

Our paper makes four key contributions to the XAI field:
\begin{enumerate}[leftmargin=2em]
\item We propose the first formal, multidimensional definition of contestability in AI systems.
\item We introduce a taxonomy of contestability dimensions and criterion clusters that practitioners can use to ensure comprehensive coverage of contestation requirements.
\item We present a design framework outlining both by-design and post-hoc mechanisms, offering concrete workflows for embedding contestability throughout the AI lifecycle.
\item We develop the \emph{Contestability Assessment Scale}, a composite metric based on over twenty weighted criteria that enables quantitative evaluation of AI systems' contestability.
\end{enumerate}

The remainder of this paper is organized as follows: Section~\ref{sec:related_work} reviews related work on contestability in AI and explainable systems, identifying key gaps in current practice. Section~\ref{sec:definition} introduces a formal, multidimensional definition of contestability and provides its mathematical formalism. Section~\ref{sec:taxonomy} presents our taxonomy and criteria for contestability, structured across human-centered, technical, legal, and organizational dimensions. Section~\ref{sec:cas} introduces the Contestability Assessment Score, a quantitative metric for evaluating AI systems. Section~\ref{sec:case_studies} applies our framework in case studies across high-, medium-, and low-risk domains. Section~\ref{sec:discussion} discusses the main findings of our work, while Section \ref{sec:limitations} presents the main limitations. We conclude in Section~\ref{sec:conlusion} with a discussion of future research directions, emphasizing the need for domain-specific contestability frameworks that ensure fairness, actionability, and feasibility.

\section{Related Work}
\label{sec:related_work}

\noindent
\paragraph{Contestability in AI Systems.} Contestability in AI has emerged as a critical mechanism for ensuring that automated decisions remain open to challenge, revision, and human oversight \cite{alharbi2024misfitting, freedman2025argumentative}. It is defined as a dynamic process enabling users to interact with AI systems, challenge decisions, and obtain revised outcomes, thereby embedding accountability and responsiveness \cite{alharbi2024misfitting, maxwell2023meaningful, almada2019human, hildebrandt2022qualification, fanni2023enhancing}. This process empowers stakeholders to question or appeal AI-generated outcomes, reinforcing system adaptability and human agency \cite{freedman2025argumentative, maxwell2023meaningful, guzey2022context}. As AI is increasingly deployed in high-stakes domains, contestability is recognized as a safeguard for users’ rights, dignity, and autonomy \cite{walmsley2021artificial, sreedharan2020bridging, srivastava2024ai, ganapati2024public, marengo2023challenges, maxwell2023meaningful, almada2019human, arif2022towards}. The literature links contestability to procedural justice and advocates for “contestability by design,” embedding mechanisms for user interaction, error correction, and critique from the outset \cite{narayanan2023risk, almada2019human, dobbe2021hard, mulligan2019shaping, hirsch2017designing}. In democratic societies, contestability upholds human dignity and autonomy by ensuring individuals can challenge impactful decisions, whether made by humans or algorithms \cite{almada2019human, kaminski2018binary}.

\noindent
\paragraph{Contestability in Explainable AI.} Explainability is widely recognized as a prerequisite for meaningful contestability, enabling users to understand, question, and act upon AI outputs \cite{aler2020contestable, freedman2025argumentative, leofante2024contestable, hildebrandt2022qualification, raees2024explainable, yurrita2023generating, guzey2022context}. Explanations should go beyond post-hoc rationalizations to support users’ rights to inquiry and redress \cite{alfrink2023contestable, raees2024explainable, lyons2021conceptualising}. Scholars distinguish between descriptive and justificatory explanations, noting that only the latter fully support contestation by addressing the “why” behind a decision \cite{maxwell2023meaningful, yurrita2023generating, lyons2021conceptualising}. Effective contestability requires explanations tailored to user goals and contexts, and embedded in interactive, actionable systems \cite{alfrink2023contestable, leofante2024contestable, maxwell2023meaningful, calem2024intelligent, raees2024explainable}. Thus, explainability and contestability are co-dependent: transparent, user-centered systems empower stakeholders to exercise informed opposition and foster trust \cite{aler2020contestable}.

\noindent
\paragraph{Current Limitations.}
Despite growing recognition, practical barriers persist: individuals often face insufficient notice, opaque explanations, and unclear appeal pathways \cite{now2018litigating, richardson2019litigating, karusala2024understanding}. The landscape remains fragmented, lacking a comprehensive framework that addresses contestability across technical, legal, and socio-ethical levels \cite{rodrigues2020legal, eubanks2018automating, benjamin2019assessing, chouldechova2018frontiers, myers2018censored, gorwa2020algorithmic}. Legal frameworks such as the EU’s GDPR enshrine rights to contestation but often lack operational clarity \cite{mendoza2017right, almada2019human, pi2024empowering, sanderson2022towards}. The notion of recourse—allowing users to alter outcomes—captures only a subset of contestability, often omitting upstream elements like explanation and process transparency \cite{ustun2019actionable, venkatasubramanian2020philosophical, almada2019human}. Existing work is often domain-specific and rarely integrated into a cross-sectoral view \cite{tomsett2018interpretable}. Our work addresses these gaps by empirically examining contestability across domains and proposing an integrated framework that formalizes the relationship between explainability and contestability, contributing actionable design principles for transparent and fair algorithmic decision-making.



\section{Defining Contestability}
\label{sec:definition}

To derive a standardized definition of contestability, we conducted a semi-automated systematic literature review following the PRISMA methodology \cite{page2021prisma}, supported by large language models (LLMs) for search expansion and thematic synthesis. Details of the review process are provided in Appendix~\ref{app:slr}. This section presents our analysis of existing definitions of contestability in the context of XAI, along with our proposed standardized definition and accompanying mathematical formalism.

\subsection{Conceptual Definition}

We define \emph{contestability} as a multidimensional, dynamic property of AI systems that enables affected stakeholders to actively challenge, scrutinize, and influence either the immediate outcomes or future decision-making processes of these systems. Contestability stands apart from explainability in its focus on agency and action rather than mere understanding.

Contestability serves three fundamental purposes in AI governance:

\begin{enumerate}[leftmargin=2em]
    \item Enabling recourse: Provides mechanisms for affected parties to dispute specific outcomes when they believe errors, biases, or unfairness have occurred.
    \item Facilitating risk analysis: By analyzing contestation patterns, organizations can identify blind spots in model design, deployment constraints, and unintended consequences.
    \item Ensuring accountability: Establishes clear pathways for human intervention and oversight, creating lines of responsibility for AI operations and impacts.
\end{enumerate}

The relationship between explainability and contestability is fundamental: while explanations provide the necessary foundation for understanding AI decisions, contestability transforms this understanding into actionable agency. Without explainability, contestation becomes arbitrary; without contestability, explanations remain passive information.

Contestability can be instantiated in two modes: \emph{by design}, when the underlying model is transparent and contestation is integrated into the system architecture; and \emph{post hoc}, when the model is opaque and contestation relies on external mechanisms such as approximate explanations, audits, or appeals.

\subsection{Mathematical Formalism}

Let $\mathcal{A}$ be an AI system producing decisions $\mathcal{D}$ over input space $\mathcal{X}$. Let $S$ be the set of stakeholders, each with capabilities $\mathcal{P}(s)$, and let $\mathcal{E}: \mathcal{D} \rightarrow \mathcal{R}$ be an explanation function mapping decisions to human-interpretable representations $\mathcal{R}$. Let $\mathcal{C}$ be the contestation space defined by the set of actions available to challenge or influence system outcomes, then contestability consists of 

\paragraph{Explanation-Level Contestability (XLC):}
\[
\text{Contest}_{\text{XLC}}(\mathcal{E}) \iff \forall d \in \mathcal{D},\, \forall s \in S,\, \exists r = \mathcal{E}(d) \in \mathcal{R}:\ r \rightarrow c \in \mathcal{C},\, \text{given}\ \mathcal{P}(s).
\]
That is, explanations must be actionable relative to stakeholder capabilities.

\paragraph{System-Level Contestability (SLC)} is define by
\[
\text{Contest}_{\text{SLC}}(\mathcal{A}) \iff \exists\, \mathcal{E}, \mathcal{C}\ \text{s.t.}\ \forall s \in S,\, \exists x \in \mathcal{X},\, d = \mathcal{A}(x),\, r = \mathcal{E}(d),\, c \in \mathcal{C}:\ \phi(c, d, \mathcal{A}) = \text{True},
\]
where $\phi$ captures successful contestation—either through immediate correction of $d$ or systemic adaptation of $\mathcal{A}$ (e.g., model update, data revision, or audit trigger).

\paragraph{Aggregate Contestability}
\[
\text{Contest}(\mathcal{A}) = \alpha\, \text{Contest}_{\text{XLC}}(\mathcal{E}) + \beta\, \text{Contest}_{\text{SLC}}(\mathcal{A}) + \gamma\, \min_{s \in S} \text{SR}(\mathcal{A}, s),
\]
where $\text{SR}(\mathcal{A}, s)$ is the contestation success rate for stakeholder $s$, and $\alpha$, $\beta$, $\gamma$ are weighting coefficients ensuring both technical validity and equity.

\subsection{Contestability Modes: By Design vs Post-Hoc}

Contestability operates through two complementary modes that reflect the level of access to the underlying predictive model:

\textbf{By Design Contestability ($\mathcal{C}_{\text{design}}$)}  
arises when the AI system is inherently transparent or modular. Stakeholders can inspect model logic, enabling rich, granular explanations and direct contestation mechanisms. Typical features include: access to model internals (e.g., feature weights, rule sets), interpretable architectures supporting fine-grained audits,
    explanation–recourse coupling via editable decision components, configurable logging and contestation workflows embedded at design time, etc.

\textbf{Post-Hoc Contestability ($\mathcal{C}_{\text{post}}$)}  
applies to black-box systems where internal logic is opaque or inaccessible. Contestation relies on external mechanisms and indirect inference. Common strategies include: post-hoc explanations (e.g., SHAP, LIME) without model transparency, regulatory appeals and legal audits based on outcomes, retrospective fairness or error pattern analysis, external complaint and feedback channels for redress.

Formally, the contestation space combines both:
\[
\mathcal{C} = \mathcal{C}_{\text{design}} \cup \mathcal{C}_{\text{post}}.
\]

While by-design mechanisms offer immediacy and efficiency, post-hoc tools ensure accountability in systems lacking built-in contestation, or where unforeseen harms emerge.

\subsection{Context Sensitivity and Limitations}
\label{sec:sensitivity}

Contestability is domain-sensitive and constrained by several context-specific variables:

\begin{enumerate}[leftmargin=2em]
    \item Latency Constraints ($\tau$): Systems operating faster than human reaction time (e.g., trading bots, autonomous weapons) preclude real-time challenges.
    \item Opacity Constraints ($\Omega$): Proprietary models or trade secrets may hinder transparency and actionable recourse.
    \item Capability Disparities ($\Delta$): Resource asymmetries (expertise, access, time) can marginalize less-resourced stakeholders.
    \item Adaptivity Constraints ($\Gamma$): Individual grievances may not address systemic harms without aggregation mechanisms.
\end{enumerate}

These factors define the contestability context:
\[
\Theta = \{\tau, \Omega, \Delta, \Gamma, \dots\}.
\]

and it has the advantage of being able to be extended to any other factors such as \emph{regulatory requirements}, \emph{application criticality}, etc. Moreover, contestability is bounded by the limitations of explainability; yet, even when full explanations are unavailable, partial contestability may arise from outcome-based signals or pattern-based audits.

\section{Contestability Taxonomy and Criteria}
\label{sec:taxonomy}

To move from abstract principles to concrete practice, we decompose contestability into four orthogonal \emph{dimensions} (Human‐Centered, Technical, Legal, and Organizational) and organize over twenty specific \emph{criteria} into five thematic clusters.  The dimensions capture the interdisciplinary nature of contestability (from user experience to regulatory compliance), while the clusters structure criteria from foundational access through process design to lifecycle integration.  This taxonomy provides a clear, systematic lens for both evaluating existing AI systems and guiding the design of new mechanisms that genuinely enable stakeholders to challenge, review, and reshape automated decisions.

\subsection{Taxonomy of Contestability Dimensions}

In our framework, each contestability criterion is classified under one of four dimensions (Human‐Centered, Technical, Legal, or Organizational) to denote its primary area of focus. This classification highlights the interdisciplinary scope of contestability and offers a structured basis for both evaluating AI systems and designing targeted mechanisms that span user experience, system architecture, regulatory compliance, and governance practices.

\begin{enumerate}[leftmargin=2em]

    \item \taghumandef dimensions refer to design and evaluation principles that prioritize the experiences, needs, and rights of individuals and communities affected by AI systems. They emphasize usability, inclusivity, emotional well-being, cultural adaptation, and participatory design. These considerations are rooted in fields such as human-computer interaction (HCI), cognitive ergonomics, and user-centered ethics.

    \item \tagtechdef dimensions capture the computational, architectural, and algorithmic properties that operationalize contestability. These include mechanisms for logging, explainability, real-time override, system integrity, and data provenance. This dimension draws from software engineering, computer security, and explainable AI.

    \item \taglegaldef dimensions reflect normative, statutory, and regulatory conditions that enshrine contestability as a right. This includes principles such as due process, privacy protection, and the enforceability of redress mechanisms under data protection laws (e.g., GDPR), administrative law, and emerging AI-specific regulations.

    \item \tagorgdef dimensions relate to institutional protocols and governance structures that embed contestability throughout the AI lifecycle. These include clearly assigned roles, escalation pathways, transparency reporting, and processes for organizational learning from contestation. This dimension is informed by institutional theory, risk management practices, and socio-technical systems design.
    
\end{enumerate}

\subsection{Contestability Criteria}

To structure the operationalization of contestability, we define a set of criteria grouped into five thematic clusters. Each cluster reflects a distinct dimension of how contestability can be embedded into AI systems—ranging from foundational access to lifecycle adaptivity. The complete list of criteria and definitions is provided in Appendix~\ref{app:criteria}.

\begin{enumerate}[leftmargin=2em]
    \item Structural Preconditions for Contestability: This cluster includes requirements that ensure contestation mechanisms are accessible, discoverable, and usable by all affected stakeholders, including those from vulnerable or underrepresented groups. It covers aspects such as resource equity, multilingual support, legal awareness, and advocacy access \taghuman\taglegal.

    \item Process Integrity and Contestation Design: This cluster focuses on the quality and fairness of contestation processes. It emphasizes actionable explanations, the ability to challenge decisions at multiple levels, and the presence of timely interventions and safeguards against retaliation \tagtech\taglegal.

    \item Governance, Traceability, and Accountability: These criteria ensure that contestation is not only available but also governed in a transparent and auditable manner. They include requirements for secure logging, regulatory oversight, role clarity, and clear boundaries on what can be contested \tagtech\taglegal\tagorg.

    \item Adaptation, Reflexivity, and Lifecycle Integration: This cluster addresses the dynamic evolution of contestability mechanisms. It supports feedback loops, barrier monitoring, co-design with users, and proportional safeguards based on system risk \tagorg\taghuman\tagtech.

    \item Supportive Infrastructure and Normative Commitments: The final cluster includes ethical and technical underpinnings such as privacy-preserving mechanisms, system resilience, and transparent reporting of contestation outcomes \tagtech\tagorg.

\end{enumerate}

These clusters provide the basis for evaluating and implementing contestability in AI systems.  Appendix \ref{app:taxonomy} presents a detailed mapping of these criteria onto diverse scenarios, categorized by their associated AI risk and degree of reliance on AI systems.

\section{The Contestability Assessment Score (CAS)}
\label{sec:cas}

Based on the proposed contestability taxonomy, we can define a formal, quantifiable framework for contestability
and a systematic evaluation of contestability in AI systems, we introduce a formal definition grounded in system-level properties that contribute to the feasibility of contestation. 
Building on this foundation, we propose the \emph{Contestability Assessment Score (CAS)} — a composite, continuous metric in the range $[0, 1]$ that quantifies the extent to which a system enables stakeholders to challenge, influence, and seek redress over automated decisions. 
CAS integrates multiple dimensions reflecting both design-time and runtime affordances for contestability, with higher values indicating stronger support for transparency, accessibility, responsiveness, and human oversight. 

The CAS is derived from a structured self-assessment questionnaire (Appendix~\ref{app:questionnaire}), organized around eight properties:

\begin{enumerate}[leftmargin=2em]
  \item \textbf{Explainability}:  
  Evaluates the system's transparency.  
  0 = No explanations; 1 = Post-hoc or approximated explanations; 2 = Intrinsically explainable model.
  
  \item \textbf{Openness to Contestation}:  
  Measures who has access to contestation pathways.  
  0 = No contestation; 1 = Expert-only; 2 = Broad stakeholder access.

  \item \textbf{Traceability}:  
  Aggregates five subcomponents (e.g., auditability, granularity, retention, transparency, and error tracking), each scored from 0 to 2.

  \item \textbf{Built-in Safeguards}:  
  Captures the presence of mechanisms preventing retaliation or harm during contestation.  
  0 = None; 1 = Present.

  \item \textbf{Adaptivity}:  
  Assesses whether the system updates based on contestation events.  
  0 = Static; 1 = Reactive adjustments; 2 = Proactive continuous learning.

  \item \textbf{Auditing Mechanisms}:  
  Reflects the scope and independence of contestability audits.  
  0 = None; 1 = Internal audit; 2 = Independent external audit.

  \item \textbf{Ease of Contestation}:  
  Evaluates practical accessibility via 10 binary criteria (e.g., multilingual access, fee waivers, assistance availability, defined resolution timelines).

  \item \textbf{Explanation Quality}:  
  Computed using the System Causability Scale (SCS) based on user-rated agreement (10 items, 1--5 Likert scale), yielding a maximum of 50 points.
\end{enumerate}

Let $P = 8$ be the number of AI system's properties related to contestability and let $s_p \in S_p$ be the raw score for the $p-th$ property, where $p = 1, \cdots, P$ and $S_p$ is the set of all possible raw score values.
Then the \emph{CAS} is defined as a linear combination of the properties' raw scores: 

\[
\mathrm{CAS} = \sum_{p=1}^{P} \lambda_p \cdot s_p \cdot n_p  \qquad \text{with } \sum_p \lambda_p = 1.
\]

The normalization term is given by $n_p = 1/s_p^{max}$, with $s_p^{max}$ being the maximum possible value assumed by the raw score.
To each property $p$ is assigned a weight coefficient $\lambda_p$ reflecting not only its intrinsic theoretical and empirical significance, but also its dependencies on other properties arising from logical or functional "consequentiality." 
For instance, $p_2=$ \emph{openness to contestation} may rely on the system’s $p_1=$ \emph{explainability}, and $p_7=$ \emph{ease of contestation} in turn builds upon the former. 
To account for such ``inter-dependencies", each property was assigned a priority level, with critical properties receiving higher weights.
\[
1 > \lambda_1 > \lambda_2 = \lambda_3 = \lambda_4 > \lambda_5 = \lambda_6 > \lambda_7 = \lambda_8 > 0
\]

Given that the weights should sum to 1, among the acceptable configurations, the following set of weights was selected as the final allocation:

\[
\overbrace{0.30}^{\lambda_1} + \overbrace{0.12 + 0.12 + 0.12}^{\lambda_{2,3,4}} + \overbrace{0.10 + 0.10}^{\lambda_{5,6}} + \overbrace{0.07 + 0.07}^{\lambda_{7,8}} = 1
\]


The strength of the Contestability Assessment Score lies in its versatility: the proposed formalization considers solely eight known system's properties that support contestability at various levels, each one of them evaluated according to specific criteria. However, there is nothing to prevent the inclusion of additional properties that more precisely describe the contestable system or select different values for the weight parameters, while respecting the priorities. In this way, it is possible to indirectly estimate the level of contestability of a system through its properties, whose evaluation criteria are known and directly assessable.

\section{Case Studies}
\label{sec:case_studies}

To illustrate the practical relevance and diagnostic capabilities of our contestability framework, we consider three AI systems with increasing levels of risk exposure:
\begin{enumerate}[leftmargin=2em]
    \item \emph{High-risk}: Automated Radiology Diagnosis as an example of AI system in healthcare;
    \item \emph{Medium-risk}: Automated Credit Scoring for loan application; 
    \item \emph{Low-risk}: Personalized News Recommendation system.
\end{enumerate}

For each case study, we assess contestability using our proposed taxonomy and compute the Contestability Assessment Score, highlighting how system properties enable—or hinder—stakeholder challenge, influence, and redress. 
Only the high-risk example will be entirely presented in the paper, while the other case studies are available in the Appendix \ref{app:case_study}.

\subsection{Case Study 1: High Risk - AI in Healthcare}

The \emph{Automated Radiology Diagnosis} system is a deep learning system deployed in a hospital to assist radiologists in diagnosing chest X-rays. It provides a binary prediction (disease/no disease) and a heatmap highlighting salient image regions.

\paragraph{Contestability Criteria:} 

\begin{enumerate}[leftmargin=2em]

    \item{\textbf{\taghumannop}} Explanations are provided as heatmaps, but patients and non-expert clinicians find them difficult to interpret. No multilingual or accessible contestation channels exist for patients.
    
    \item{\textbf{\tagtechnop}} The system logs predictions and explanations, but logs are only accessible to IT staff. No real-time override is available; only retrospective review.
    
    \item{\textbf{\taglegalnop}} Patients are not proactively informed of their right to contest, and there is no formal appeal process. Regulatory audits are internal.
    
    \item{\textbf{\tagorgnop}} Escalation pathways exist for clinicians but not for patients. Feedback from contestations is not systematically used to improve the model.

\end{enumerate}

\paragraph{Contestability Assessment Score:}

When assessed using the proposed CAS, the system’s limitations become quantifiable. By qualitatively evaluating the system, we can attribute a score of 1 for \emph{explainability}, reflecting its reliance on post-hoc explanations. \emph{Openness to contestation} is similarly \emph{limited}, with a score of 1, as only clinicians—not patients—can initiate a challenge. \emph{Traceability} is partial, scoring 6 out of 10, due to the restricted access to logs. \emph{Built-in safeguards} are minimal, earning a score of 1, and the \emph{system’s adaptivity} is limited to reactive updates, also scoring 1. Auditing is internal only, leading to a low score of 1. The \emph{ease of contestation} is very low, at 2 out of 10, as escalation is possible only for clinicians and not for patients. Finally, \emph{the quality of the explanations, as rated by the users, is mixed, with a score of 25 out of 50}, reflecting the limited contestability capabilities of the system. Each score has to be normalized by the proper maximal value.

\input{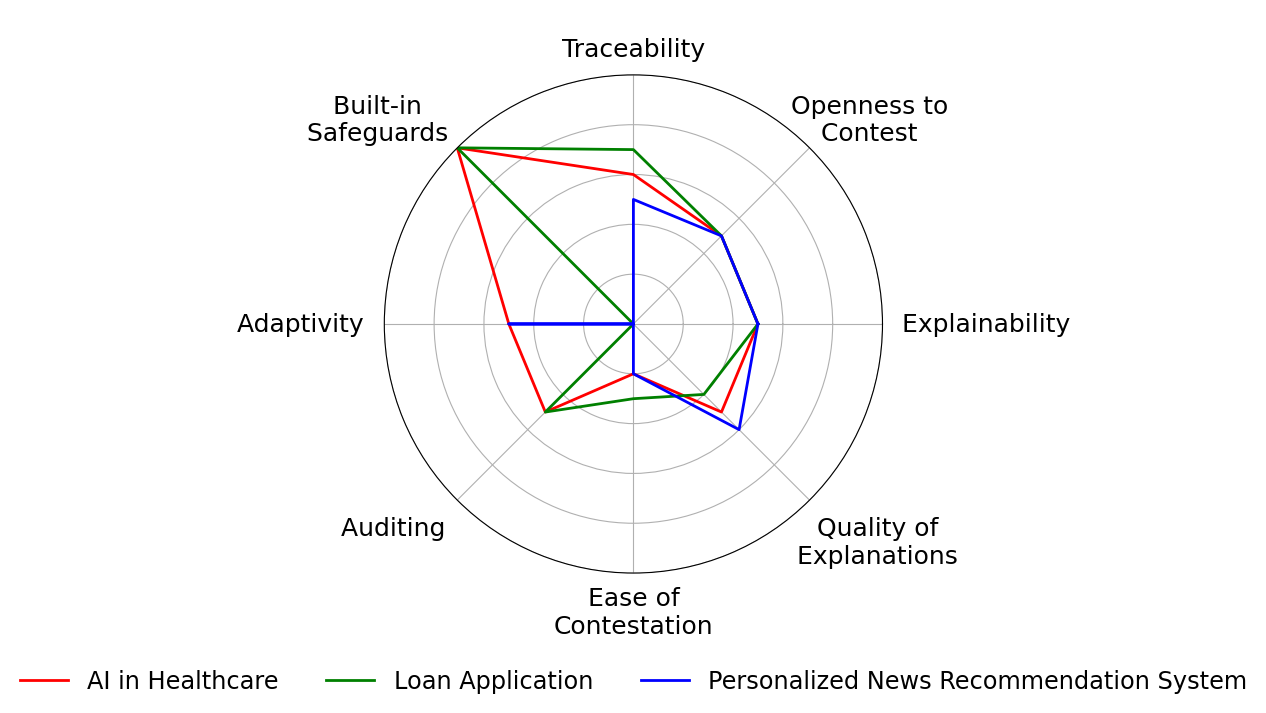}

\paragraph{Recommendations:}
Based on the CAS assessment and considering this application's categorization as High AI Reliance and High Risk (Appendix \ref{app:taxonomy}), the following \emph{highly} or \emph{moderately} feasible recommendations are proposed to enhance compliance with contestability safeguards.

\begin{enumerate}[leftmargin=2em]

    \item \textbf{\taghumannop} Patient education (highly feasible) involves informing patients of their rights. Accessible explanations and multilingual channels (moderately feasible) require significant resources for effective communication.

    \item \textbf{\tagtechnop} Retrospective review (highly feasible) allows post-decision challenges. Expanding log access (moderately feasible) balances transparency with privacy. Improving explanation quality (highly feasible) uses feedback for better understanding.

    \item \textbf{\taglegalnop} Formal appeals and clear boundaries (highly feasible) establish a process and its limits. External auditing (moderately feasible) requires resources and privacy safeguards.

    \item \textbf{\tagorgnop} Feedback loops and clear roles (highly feasible) integrate contestation into system improvement. Stakeholder co-design (moderately feasible) demands a commitment to participatory methods.

\end{enumerate}

The revised CAS for the updated system would be: 0.622 (if only highly feasible changes are implemented), 0.927 (if only the moderately feasible recommendations are implemented). In the table \ref{tab:case-study-1} the improvements for each property are highlighted by color.

\input{tables/cas_clinical}

\section{Discussion}\label{sec:discussion}
This section presents the key scientific conclusions derived from our investigation of contestability in XAI, highlighting their implications for the design, evaluation, and governance of AI systems.

\noindent
\textbf{Finding 1: Contestability as a Distinct Construct in XAI.}
Our work establishes contestability as a distinct and measurable construct within XAI, differentiated from related concepts like explainability and recourse. We provide a rigorous, multidimensional definition of contestability as the actionable capacity for stakeholders to challenge, influence, and seek redress over AI-driven decisions. This definition provides a theoretical foundation for future research and enables the formal analysis of contestability mechanisms.

\noindent
\textbf{Finding 2: An Interdisciplinary Framework for Operationalizing Contestability.}
We present a novel, interdisciplinary framework for operationalizing contestability, comprising a taxonomy of dimensions (human-centered, technical, legal, organizational) and actionable criteria. This framework provides a structured approach for designing and evaluating contestability mechanisms, ensuring that they are not only technically feasible but also ethically sound and socially robust. The framework's modularity allows for adaptation to diverse application domains and regulatory contexts.

\noindent
\textbf{Finding 3: The Contestability Assessment Score as a Quantitative Metric for System Evaluation.}
We introduce the Contestability Assessment Score as a quantitative metric for evaluating the degree to which AI systems support contestation. Our case studies demonstrate that CAS can effectively diagnose shortcomings in existing systems and guide targeted improvements. The CAS provides a valuable tool for researchers and practitioners seeking to benchmark and compare the contestability of different AI systems.

\noindent
\textbf{Finding 4: Actionable Tools for Aligning AI Systems with Regulatory Mandates.}
A key finding of this work is the demonstration that actionable tools—specifically, a clear definition, a comprehensive framework, and a quantitative assessment metric—can directly align AI systems with emerging regulatory mandates. By providing stakeholders with the resources to translate abstract regulatory principles into concrete system properties, we facilitate the development of more accountable, participatory, and legally robust AI systems. This represents a significant step towards bridging the gap between AI governance and practical implementation.

\noindent
\textbf{Finding 5: Societal Impact and the Need for Responsible Implementation.}
Our proposed framework for contestability in XAI has significant potential for positive societal impact, promoting accountability, fairness, and procedural justice, particularly in high-stakes domains. However, we also acknowledge potential risks: contestability mechanisms may be misused, may exacerbate power asymmetries, or may impose undue burdens on stakeholders. Therefore, future implementations must prioritize usability, transparency, and equitable access, and should be evaluated in context-sensitive ways that reflect diverse sociotechnical environments. This underscores the need for responsible design and deployment to ensure that contestability mechanisms empower stakeholders and mitigate algorithmic harms.


\section{Limitations} \label{sec:limitations}

Our work provides a foundation for contestability in XAI, but several limitations warrant consideration. First, the CAS framework relies on subjective assessments and may not generalize across all domains. Future work should explore automated assessment and participatory weighting. Second, our framework does not fully address contestability in systems with extremely fast predictions, requiring alternative mechanisms like ex-post audits. Third, contestability requirements will evolve, necessitating ongoing updates to our framework. Despite these limitations, our work offers a starting point for designing more accountable and responsive AI systems, stimulating further innovation in responsible AI.

\section{Conclusion and Future Work}\label{sec:conlusion}

This paper is the first to address the persistent gap between regulatory mandates and practical implementations of contestability in explainable AI. We presented a formal, multidimensional definition of contestability grounded in both theoretical and applied perspectives, distinguishing it from explainability by its emphasis on stakeholder agency, actionable recourse, and systemic oversight. Our modular framework synthesizes contestability mechanisms across human-centered, technical, legal, and organizational dimensions, and our Contestability Assessment Score operationalizes these concepts into a quantitative metric. Through diverse case studies, we demonstrated how current AI systems fall short of regulatory ideals and how our framework enables concrete improvements. Future work should focus on translating this general framework into domain-specific contestability blueprints that balance actionability, fairness, and feasibility across varied stakeholder groups and application contexts. By formalizing contestability and offering tools for its assessment and implementation, this work lays the foundation for more accountable, participatory, and legally robust AI systems.

\clearpage

\bibliographystyle{unsrtnat}
\bibliography{neurips_2025}

\clearpage

\appendix\label{sec:appendix}

\input{appendix}

\end{document}

%% file: imgs/radar_chart.tex
\begin{figure}[htbp]
    \centering
    \includegraphics[width=0.8\textwidth]{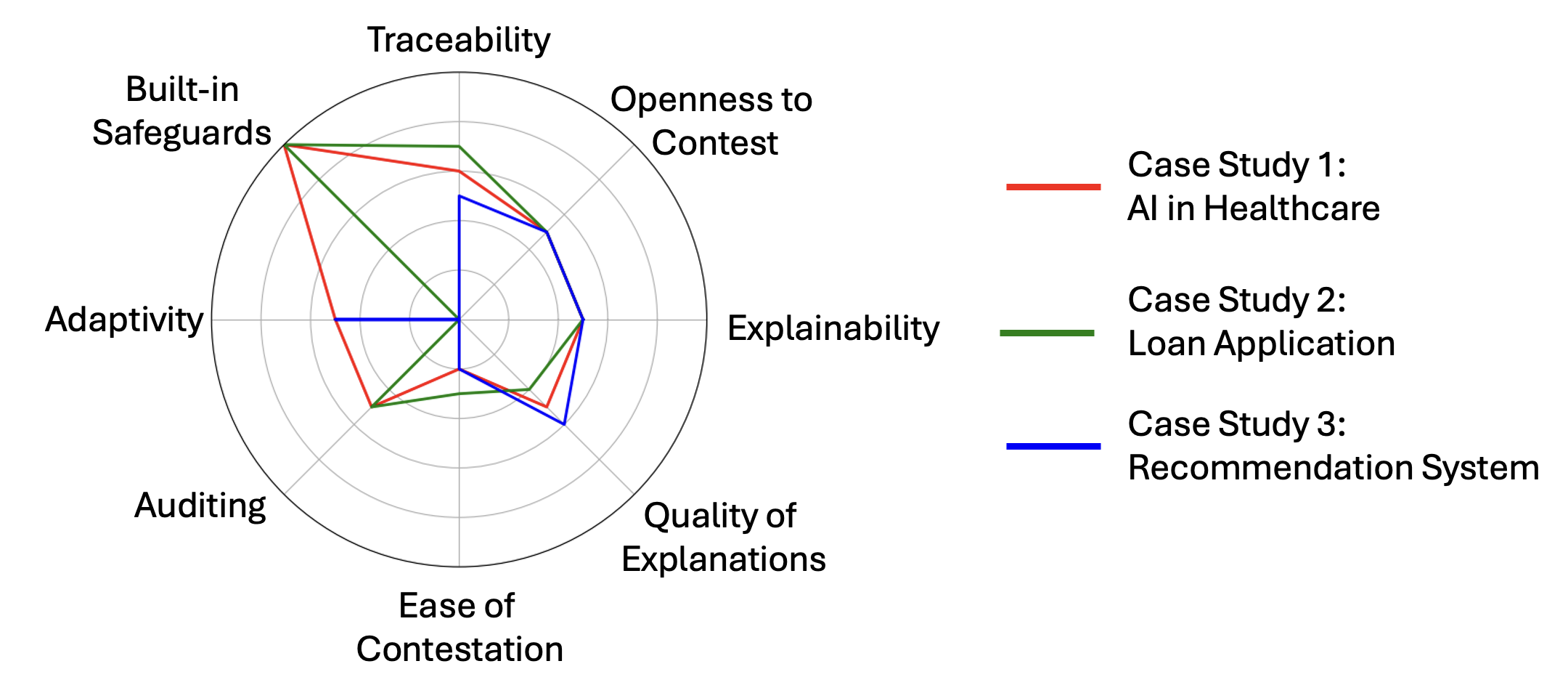}
    \caption{Contestability Assessment Score for AI Systems grouped by category. Detailed results for the finance and recommendation cases are provided in Appendix~\ref{app:case_study}.}
    \label{fig:radar_chart}
\end{figure}






%% file: tables/cas_clinical.tex
\begin{table}[!h]
\centering
\caption{Case Study 1: High AI-Reliance / High-Risk. Comparison of CAS Scores Across Different Implementations of Contestability Scenarios: HF corresponds to \textit{highly feasible} changes, while MF is \textit{moderately feasible}. }
\resizebox{\columnwidth}{!}{
\begin{tabular}{lccccccccc}\label{tab:case-study-1}
\textbf{Property} & \textbf{Max} & \textbf{Weight} & \textbf{Score} & \textbf{CAS System} & \textbf{ScoreHF} & \textbf{CAS HF} & \textbf{Score MF} & \textbf{CAS MF} \\
\hline
Explainability & 2 & 0.30 & 1 & 0.150 & 1 & 0.150 & \cellcolor[HTML]{EAFAF1}2 & \cellcolor[HTML]{EAFAF1}0.300 \\
\hline
Openness to Contestation & 2 & 0.12 & 1 & 0.060 & 1 & 0.06 & 1 & 0.120 \\
\hline
Traceability & 10 & 0.12 & 6 & 0.072 & 6 & 0.072 & \cellcolor[HTML]{EAFAF1}8 & \cellcolor[HTML]{EAFAF1}0.096 \\
\hline
Built-in Safeguards & 1 & 0.12 & 1 & 0.120 & 1 & 0.120 & 1 & 0.120 \\
\hline
Adaptivity & 2 & 0.10 & 1 & 0.050 & \cellcolor[HTML]{EAF2FA}2 & \cellcolor[HTML]{EAF2FA}0.100 & 2 & 0.100 \\
\hline
Auditing & 2 & 0.10 & 1 & 0.050 & 1 & 0.050 & \cellcolor[HTML]{EAFAF1}2 & \cellcolor[HTML]{EAFAF1}0.100 \\
\hline
Ease of Contestation & 10 & 0.07 & 2 & 0.014 & \cellcolor[HTML]{EAF2FA}5 & \cellcolor[HTML]{EAF2FA}0.035 & \cellcolor[HTML]{EAFAF1}6 & \cellcolor[HTML]{EAFAF1}0.042 \\
\hline
Explanation Quality & 50 & 0.07 & 25 & 0.035 & 25 & 0.035 & \cellcolor[HTML]{EAFAF1}35 & \cellcolor[HTML]{EAFAF1}0.049 \\
\hline
\hline
\textbf{Total CAS} &  &  &  & \textbf{0.551} &  & \textbf{0.622} &  & \textbf{0.927} \\
\hline
\end{tabular}
}
\end{table}

%% file: appendix.tex
\section{Semi-Automated Systematic Literature Review}\label{app:slr}

To ensure a comprehensive and unbiased synthesis of the literature on contestability in XAI, we followed the PRISMA methodology \cite{page2021prisma}. Our aim was to develop a robust framework for contestability in XAI by systematically identifying, reviewing, and synthesizing relevant research.

\subsection{Research Questions}

Our review was guided by the following research questions:

\begin{itemize}
    \item Definition and Conceptualization: How is contestability defined and conceptualized in the context of artificial intelligence and explainable AI?

    \item What are the key properties or dimensions used to characterize contestability in AI systems?

    \item What are the existing frameworks or models for implementing contestability in AI, particularly in explainable AI systems?

    \item What are the challenges and limitations identified in the literature regarding the implementation of contestability in AI?

    \item How does contestability relate to other important concepts in AI ethics and governance, such as transparency, accountability, and fairness?

    \item What are the proposed methods or techniques for evaluating the effectiveness of contestability mechanisms in AI systems?

    \item How does contestability in AI vary across different application domains or sectors (e.g., healthcare, finance, criminal justice)?

\end{itemize}

\subsection{Search Strategy}

We constructed a search query using the following keywords:
\textit{( contestab* ) AND ("Artificial Intelligence" OR "AI" OR "explainable AI" OR "explainable artificial intelligence" OR "XAI" )}. The search was conducted across the following databases: Scopus, ACM Digital Library, PubMed, IEEE Xplore, and Arxiv. Manual searches and reference snowballing were also performed to ensure coverage of seminal works.

\subsection{Screening and Selection}

The initial search yielded 94 records. After removing duplicates and applying inclusion/exclusion criteria, 75 unique records remained. Further screening based on title and abstract led to 60 papers for full-text review. After full-text assessment, 92 papers were included in the final synthesis (including additional relevant works identified through manual review and reference mining). Table \ref{tab:lit_rev_results} presents the articles screened in each phase of the PRISMA framework.

\begin{table}[h!]
\begin{tabular}{lcccc}
\hline
\textbf{Database} & \textbf{Initial} & \textbf{After Duplicates} & \textbf{After Screening} & \textbf{Final} \\
\hline
Scopus           & 62 & 62 & 54 & 54 \\
ACM Digital Lib  & 14 & 2  & 2  & 2  \\
PubMed           & 7  & 2  & 2  & 2  \\
IEEE Xplore      & 11 & 9  & 2  & 2  \\
Arxiv            & 0  & 0  & 0  & 23 \\
Manual           & 0  & 0  & 0  & 9  \\
\textbf{Total}   & 94 & 75 & 60 & 92 \\
\hline
\end{tabular}
\caption{Number of articles retrieved, filtered, and included from different academic databases during the systematic review process}
\label{tab:lit_rev_results}
\end{table}

\subsection{Inclusion / Exclusion Criteria}

Inclusion and Exclusion Criteria
To ensure the relevance and quality of the synthesized literature, we established clear inclusion and exclusion criteria. Studies were included if they explicitly focused on contestability within the context of artificial intelligence or explainable AI, encompassing its definition, conceptualization, or key properties. We also included studies that discussed frameworks, models, or methodologies for implementing contestability in AI systems, as well as those that examined contestability in relation to broader AI ethics and governance principles, such as transparency, accountability, fairness, and user participation. Furthermore, studies that discussed evaluation methods or techniques for assessing contestability in AI systems, or analyzed sector-specific applications of contestability (e.g., healthcare, finance, criminal justice), were considered eligible. Finally, only studies published in English within the last 10 years were included, with the exception of seminal works that have significantly shaped the field.

Conversely, several types of studies were excluded. Non-peer-reviewed articles, such as opinion pieces, editorials, blog posts, and white papers, were excluded to maintain the rigor of the review. Studies that mentioned contestability only in passing, without substantial discussion or analysis, were also excluded, as were those that addressed contestability in a non-AI context. Studies that focused on AI ethics broadly, without engaging deeply with contestability as a distinct concept, were not included. Additionally, studies lacking theoretical, empirical, or methodological contributions to the understanding or implementation of contestability were excluded. Duplicate publications or multiple reports of the same study were removed to avoid redundancy. Finally, studies not published in English or published more than 10 years ago (unless they were foundational works) were excluded to ensure the review focused on current and relevant research.

\subsection{Semi-Automated LLM-Based Extraction Methodology}

To ensure both rigor and reproducibility in our literature review, we implemented a semi-automated pipeline for data extraction and synthesis. Initially, papers were manually retrieved from the selected databases and screened according to the predefined inclusion and exclusion criteria. Subsequently, we employed \textit{Mistral OCR} \footnote{\url{https://docs.mistral.ai/capabilities/document/}} to convert the PDF documents into markdown text, extracting not only the textual content but also any embedded images and tables. Mistral OCR is a high-performance Optical Character Recognition (OCR) API developed by \textit{Mistral AI}. It is designed to extract text, tables, images, and complex layouts from PDFs and images, while preserving the original document structure. Unlike traditional OCR tools, Mistral OCR outputs structured data (such as Markdown or JSON), making it especially suitable for AI-driven applications and retrieval-augmented generation (RAG) systems. It is natively multilingual, extremely fast, and excels at handling scientific, legal, and business documents with advanced formatting \cite{NEURIPS2020_6b493230}.

After applying Mistral OCR to the collected PDFs, the resulting markdown documents were then split into semantic chunks, which were embedded using \textit{OpenAI’s} embedding model and stored within a \textit{Neo4J} vector database to facilitate efficient retrieval. For each of our research questions, we defined a specific set of extraction criteria, such as the definition of contestability, key properties, existing frameworks, and evaluation methods. We then designed and iteratively refined prompts tailored to each criterion. Utilizing a RAG approach, we queried the vector database to extract relevant information from each paper using OpenAI's GPT-4o-mini. Finally, the extracted data was systematically collected and saved in CSV format, enabling further human analysis and synthesis. 

This hybrid approach enabled the analysis of a large pool of text in a shorter amount of time compared to the traditional PRISMA methodology, which requires manual processing and analysis of all papers. Given that the primary goal of this literature review was to understand the various definitions of contestability, identify key properties, and determine the domains of application, this approach is suitable and presents minimal risks. Furthermore, the risk of LLM hallucinations in RAG approaches is reduced, and traceability mechanisms were implemented in our RAG approach to track the source pages and text chunks used by the Neo4j information retrieval process to answer queries.

\clearpage

\section{Criteria for Contestability}\label{app:criteria}

After conducting the systematic literature review and analyzing its outputs, it became evident that the criteria for contestability are interconnected and mutually reinforcing. Effective contestability requires a holistic approach that considers all dimensions, from the structural preconditions that enable participation to the supportive infrastructure that ensures trustworthiness and resilience. In this section, we organize these criteria into distinct but interdependent clusters, each reflecting a critical aspect identified in the literature as essential for realizing contestability in real-world AI systems.

\subsection{Structural Preconditions for Contestability}

This cluster outlines the essential conditions that enable stakeholders to engage with, understand, and challenge AI decisions. Without these foundations, contestation remains inaccessible to many, especially vulnerable or underrepresented groups. The criteria in this cluster are:

\begin{itemize}[leftmargin=1.5em]
    \item[] \textbf{Accessibility and Resource Equity} \taghuman
    : Contestation mechanisms must be discoverable, usable, and affordable for all affected stakeholders, including those with limited digital literacy, disabilities, or socio-economic disadvantage.
    
    \item[] \textbf{Multichannel Access and Advocacy} \taghuman
    : Systems must provide diverse channels for contestation (digital, phone, in-person), including mechanisms for third-party or legal advocacy.

    \item[] \textbf{Cultural and Linguistic Adaptation} \taghuman
    : Interfaces and contestation pathways must be localized to stakeholders’ cultural, legal, and linguistic contexts.

    \item[] \textbf{Education and Right-to-Know} \taglegal
    : Users must be proactively informed of their rights to contest, how to access mechanisms, and how decisions and data are processed.
\end{itemize}

\subsection{Process Integrity and Contestation Design}

This cluster focuses on the design quality and effectiveness of contestation mechanisms. It ensures that explanations are \emph{actionable}, contestation is possible at \emph{multiple decision points}, and outcomes can be \emph{meaningfully influenced}. It also includes safeguards for \emph{timely intervention} in critical domains and protection against retaliation. Together, these criteria ensure contestability is not symbolic but \emph{operational}, \emph{fair}, and \emph{grounded in users' rights}.

\begin{itemize}[leftmargin=1.5em]
    \item[] \textbf{Explainability--Recourse Coupling} \tagtech
    : Explanations must not only clarify \emph{why} a decision was made but also show \emph{how} it can be contested or changed.

    \item[] \textbf{Granular Contestability} \tagtech
    : Stakeholders should be able to challenge not only final outcomes but also intermediate data, causal assumptions, and model parameters.

    \item[] \textbf{Actionability and Outcome Influence} \taglegal
    : Contestation must carry the possibility of outcome change (e.g., correction, override), not serve as performative transparency.

    \item[] \textbf{Real-Time Contestation for Critical Systems} \tagtech
    : High-stakes systems (e.g., healthcare, autonomous vehicles) must support immediate human override.

    \item[] \textbf{Non-Retaliation Guarantee} \taglegal
    : Stakeholders must be protected from negative consequences when contesting AI decisions.
\end{itemize}

\subsection{Governance, Traceability, and Accountability}

This cluster defines the institutional and technical structures needed to ensure contestation processes are trustworthy, transparent, and enforceable. It emphasizes secure record-keeping, clear assignment of responsibilities, and external oversight. Together, these elements establish the foundations for accountability and ensure that contestation is available, credible and governed fairly.

\begin{itemize}[leftmargin=1.5em]
    \item[] \textbf{Traceability and Audit Logging} \tagtech
     : All contestation activity must be securely logged and auditable to ensure accountability and support oversight.

    \item[] \textbf{Regulatory Auditability and Public Oversight} \taglegal
    : Systems must allow independent regulators to evaluate the fairness and effectiveness of contestation procedures.

    \item[] \textbf{Assigned Responsibility and Escalation Paths} \tagorg
    : Clear designation of roles for intake, evaluation, and escalation is essential for contestation workflows.

    \item[] \textbf{Clarity on Contestability Boundaries} \taglegal
    : Any limitations (e.g., for national security or IP) must be explicitly justified and communicated to users.
\end{itemize}

\subsection{Adaptation, Reflexivity, and Lifecycle Integration}

This cluster focuses on the continuous evolution of contestability mechanisms throughout the AI system’s lifecycle. It emphasizes learning from contestation events, removing barriers to participation, involving users in design, and tailoring safeguards to domain-specific risks. These criteria ensure that contestability remains effective, context-aware, and responsive to both user needs and emerging system challenges.

\begin{itemize}[leftmargin=1.5em]
    \item[] \textbf{Systemic Learning from Contestation Data} \tagorg
    : Feedback from contestations should be used to identify flaws and improve future system performance.

    \item[] \textbf{Continuous Monitoring of Contestation Barriers} \tagorg
    : Institutions must regularly test and address usability, legal, or linguistic obstacles that hinder contestation.
    
    \item[] \textbf{Stakeholder Co-Design and Participatory Development} \taghuman
    : Mechanisms should be co-designed with end users and affected communities.

    \item[] \textbf{Context Sensitivity and Risk Scaling} \tagtech
    : Safeguards should be proportionate to the risk level of decisions and the dependency on AI.

    \item[] \textbf{Outcome Impact Assessment} \taghuman
    : Users should be informed of how a successful contestation would impact their access to services or benefits.
\end{itemize}

\subsection{Supportive Infrastructure and Normative Commitments}

This cluster captures the foundational supports and ethical safeguards necessary to sustain meaningful contestation. It includes technical protections such as privacy and security, as well as human-centered considerations like transparency. These elements ensure that contestation mechanisms are trustworthy, resilient, and respectful of users' rights.

\begin{itemize}[leftmargin=1.5em]
    \item[] \textbf{Privacy-Preserving Contestation} \tagtech
    : Mechanisms must protect the identity and sensitive information of the contesting party.

    \item[] \textbf{Technological Robustness and Security} \tagtech
    : Infrastructure should resist tampering, spam, or denial-of-service attacks on contestation channels.

    \item[] \textbf{Transparency of Contestation Outcomes} \tagorg
    : Aggregated contestation data (e.g., success rates, reversals) should be published with appropriate privacy safeguards.
\end{itemize}

\clearpage
\section{Contestability Taxonomy}\label{app:taxonomy}

To facilitate the practical application of contestability in AI systems, we introduce a taxonomy that systematically cross-references the level of AI reliance with the degree of contestability required. This taxonomy is designed to guide practitioners, policymakers, and system designers in selecting contestability criteria that are proportionate to both the technical role of AI and the associated risk profile.

\subsection{Taxonomy Structure and Cascading Logic}

The taxonomy is organized as a matrix, with AI reliance (Low, Medium, High) as rows and contestability level (Low, Medium, High) as columns. Each cell in the matrix represents a prototypical scenario, characterized by a set of contestability criteria and illustrative application domains.

\textbf{AI reliance} is defined as follows:
\begin{itemize}
    \item \textbf{\textcolor{DarkRed}{Low AI Reliance:}} Tasks that humans could perform without advanced computation, such as basic scheduling or simple data sorting.

    \item \textbf{\textcolor{DarkYellow}{Medium AI Reliance:}} Tasks that benefit from computational assistance or automation but remain interpretable and manageable by humans, such as document analysis or recommendation systems.

    \item \textbf{\textcolor{DarkGreen}{High AI Reliance:}} Tasks that are infeasible for humans to perform at scale or speed without AI, such as real-time image generation, large-scale data mining, or autonomous decision-making.
\end{itemize}

\noindent
\textbf{Cascading logic:}
Within each AI reliance level, contestability requirements are cumulative: a system classified as “Medium Contestability” must also satisfy the requirements for “Low Contestability” at the same AI reliance level. However, contestability criteria do not cascade across AI reliance levels; each row is independent and tailored to the technical and risk context of the system. Note that “high contestability” in a low-reliance system may require fewer or less complex mechanisms than in a high-reliance, high-stakes system.

\subsection{Taxonomy Matrix: Scenarios and Properties}

\begin{figure}
    \centering
    \includegraphics[width=\linewidth]{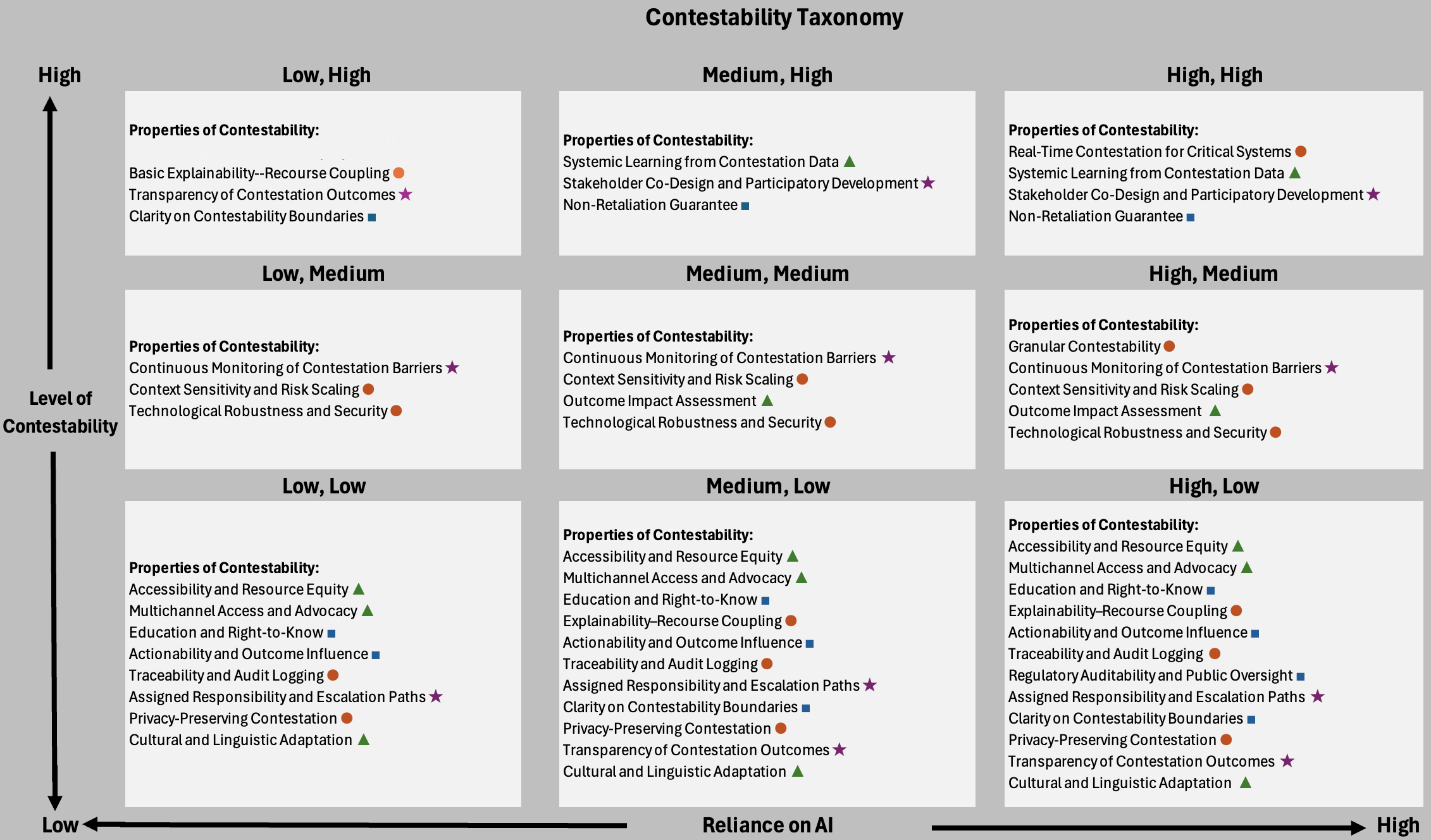}
    \caption{A visual framework mapping AI reliance and contestability level, illustrating the expected contestability properties for each scenario to guide proportionate safeguards in AI system design. The symbols correspond to the different dimensions. The triangle corresponds to \taghuman, the square to the \taglegal, the circle is the \tagtech and finally the star is \tagorg. }
    \label{fig:enter-label}
\end{figure}

\noindent
\textbf{\textcolor{DarkRed}{Low AI Reliance} / \textcolor{DarkRed}{Low Contestability}}. Systems in this quadrant are designed for efficiency and convenience, where the stakes of individual decisions are minimal. Contestability mechanisms focus on speed and basic transparency, ensuring that users can understand and, if necessary, question outcomes without significant procedural overhead.
\emph{Examples: Small retail inventory systems, library book sorting, basic CRM analytics, automated scheduling tools.}

\noindent
\textbf{\textcolor{DarkRed}{Low AI Reliance} / \textcolor{DarkYellow}{Medium Contestability.}} Here, AI systems play a more active role in users’ lives, often providing expert advice or managing personal information. Contestability mechanisms expand to include clearer explanations, accessible feedback channels, and the ability to escalate concerns.
\emph{Examples: FAQ bots, AI-powered scheduling assistants, business consultant models, online education bots.}

\noindent
\textbf{\textcolor{DarkRed}{Low AI Reliance} / \textcolor{DarkGreen}{High Contestability.}} In this scenario, the potential impact of AI decisions increases, requiring robust auditability and recourse. Systems must support detailed logging, transparent decision rationales, and formal dispute processes.
\emph{Examples: Automated grading tools with formal appeals, basic HR screening tools with audit trails, customer support ticket routing with dispute resolution.}

\noindent
\textbf{\textcolor{DarkYellow}{Medium AI Reliance} / \textcolor{DarkRed}{Low Contestability.}} These systems operate largely in the background, augmenting existing processes without direct user interaction. Contestability is minimal, often limited to internal review or basic error reporting.
\emph{Examples: School plagiarism checkers, local government permit apps, donor matching for non-profits, social media training algorithms.}

\noindent
\textbf{\textcolor{DarkYellow}{Medium AI Reliance} / \textcolor{DarkYellow}{Medium Contestability.}} AI systems in this category provide actionable insights or recommendations, directly influencing user decisions. Contestability mechanisms include user-facing explanations, the ability to challenge or appeal decisions, and transparent escalation pathways.
\emph{Examples: E-commerce recommendation engines, HR resume screening, job matching platforms, legal research tools, research chatbots.}

\noindent
\textbf{\textcolor{DarkYellow}{Medium AI Reliance} / \textcolor{DarkGreen}{High Contestability.}} 
These systems are deeply integrated into decision-making workflows, often collaborating with expert users. Contestability is comprehensive, with mechanisms for real-time override, participatory feedback, and continuous improvement based on contestation data.
\emph{Examples: Credit scoring algorithms with applicant recourse, pharmaceutical discovery platforms with audit trails, radiology AI with clinician override, AI peer review for academic publishing.}

\noindent
\textbf{\textcolor{DarkGreen}{High AI Reliance} / \textcolor{DarkRed}{Low Contestability.}} Highly autonomous systems with limited transparency or recourse, often due to technical opacity or proprietary constraints. The lack of contestability poses significant risks, especially as system impact grows.
\emph{Examples: Deepfake generation tools, high-frequency trading bots, automated cybersecurity agents, content recommendation engines.}
\emph{Note: In high-reliance, high-impact domains, low contestability is generally undesirable and may be prohibited by regulation.}

\noindent
\textbf{\textcolor{DarkGreen}{High AI Reliance} / \textcolor{DarkYellow}{Medium Contestability.}} 
These systems support expert users by providing advanced analytics or triage, with contestability mechanisms that enable oversight, guided review, and limited intervention.
\emph{Examples: Radiology AI assistants with review, insurance fraud detection with appeals, university admissions triage, mortgage approval systems, social media moderation (e.g., shadow banning) with escalation.}

\noindent
\textbf{\textcolor{DarkGreen}{High AI Reliance} / \textcolor{DarkGreen}{High Contestability.}} The highest level of both AI reliance and contestability, these systems operate in high-stakes, real-time environments where errors can have catastrophic consequences. Contestability mechanisms are extensive, including real-time human override, external audits, and transparent reporting of contestation outcomes.
\emph{Examples: Clinical trial design AI, self-driving cars, AI-powered judicial sentencing aids, medical diagnostics, real-time medical triage systems.}

\subsection{Practical Implications}

This taxonomy is intended as a practical tool for system designers, regulators, and stakeholders. By mapping a system’s AI reliance and desired contestability level, practitioners can identify the minimum set of criteria required and ensure that contestability mechanisms are both proportionate and effective. The cascading structure ensures that higher contestability levels build upon foundational safeguards, while the independence of AI reliance levels allows for domain-specific tailoring.

For a detailed mapping of contestability criteria to each scenario, including recommended mechanisms and implementation examples, see Table~\ref{tab:taxonomy_matrix}. This taxonomy is adaptive and intended to evolve alongside advances in AI technology and regulatory expectations. Updates should be informed by empirical evidence, regulatory changes, and stakeholder feedback. As new risks and use cases emerge, the criteria and examples in each cell can be updated to reflect best practices and domain-specific requirements.

We invite the XAI community, industry practitioners, and regulators to adopt, critique, and extend our framework, with the goal of establishing contestability as a standard, measurable, and actionable property of responsible AI systems.

\clearpage
\section{Self-Questionnaire}\label{app:questionnaire}

\subsection{Contestability}

\textbf{Explainability}: What level of explainability does the AI system provide? 

\begin{itemize}[leftmargin=1.5em]
    \item \emph{No Explanations} (0 pts): The system provides no explanations for its decisions or actions.
    \item \emph{Approximated Explanations} (1 pt): The system provides approximated explanations - narrative-driven generative responses that mimic human explanation but lack true causal insight and have no access to the model’s internal reasoning process behind particular decision process.
    \item \emph{Post-Hoc Explanations} (1 pt): The system provides post-hoc explainability methods.
    \item \emph{Intrinsically Explainable Model} (2 pts): The system is explainable by design, providing real explanations at a user level.
\end{itemize}

\textbf{Openness to Contest}: How open is the system to contestation or debate by stakeholders? 

\begin{itemize}[leftmargin=1.5em]
    \item \emph{No Contestation} (0 pts): The system does not allow contestation.
    \item \emph{Experts-Only} (1 pt): Only experts can contest or debate the system’s decisions.
    \item \emph{Broad Stakeholder Access} (2 pts): End-users can contest the system’s decisions.
\end{itemize}

\textbf{Traceability}: What level of logging does the system provide? Evaluate the system across the following criteria, assigning points for each based on the level of implementation. Each criterion is scored as: Fully Implemented (2 pts), Partially Implemented (1 pt), or Not Implemented (0 pt). The total score for Traceability is out of 10.

\begin{itemize}[leftmargin=1.5em]
    \item Granularity of Logs:
    \begin{itemize}[leftmargin=1.5em, label=\(\cdot\)]
        \item \emph{Not Implemented} (0 pts): The system provides minimal or no logging beyond basic operational logs.
        \item \emph{Partially Implemented} (1 point): The system provides basic technical logs but lacks detail in some areas (e.g., only inputs and outputs, no transformations).
        \item \emph{Fully Implemented} (2 pts): The system provides fine-grained records of inputs, transformations, model states, and outputs.
    \end{itemize}
    \item Accessibility of Logs:
    \begin{itemize}[leftmargin=1.5em, label=\(\cdot\)]
        \item \emph{Not Implemented} (0 pts): Logs are not accessible to anyone outside the system.
        \item \emph{Partially Implemented} (1 pt): Logs are accessible only to internal teams (e.g., developers), not external stakeholders.
        \item \emph{Fully Implemented} (2 pts): Logs are accessible to relevant stakeholders (e.g., users, auditors) through a secure interface.
    \end{itemize}
    \item Retention and Audit Trail:
    \begin{itemize}[leftmargin=1.5em, label=\(\cdot\)]
        \item \emph{Not Implemented} (0 pts): Logs are not retained or are deleted immediately after use.
        \item \emph{Partially Implemented} (1 pt): Logs are retained for a short period (e.g., less than 1 month) or lack a clear audit trail.
        \item \emph{Fully Implemented} (2 pts): Logs are retained for a sufficient period (e.g., 1 year or more) and provide a clear audit trail for tracing decisions.
    \end{itemize}
    \item Transparency of Logging Process:
    \begin{itemize}[leftmargin=1.5em, label=\(\cdot\)]
        \item \emph{Not Implemented} (0 pts): No documentation or transparency about the logging process.
        \item \emph{Partially Implemented} (1 pt): The logging process is documented internally but not communicated to stakeholders.
        \item \emph{Fully Implemented} (2 pts): The system documents and communicates its logging process to stakeholders (e.g., users are informed about what is logged).
    \end{itemize}
    \item Error and Anomaly Tracking:
    \begin{itemize}[leftmargin=1.5em, label=\(\cdot\)]
        \item \emph{Not Implemented} (0 pts): No specific logging for errors or anomalies.
        \item \emph{Partially Implemented} (1 pt): The system logs errors but lacks detailed context (e.g., only error codes, no context).
        \item \emph{Fully Implemented} (2 pts): The system logs errors, anomalies, and unexpected behaviors with detailed context for debugging.
    \end{itemize}
\end{itemize}

\textbf{Built-in Safeguards}: Does your AI system  have mechanisms to prevent harmful decisions? This also ensures the system is less likely to make errors that users need to challenge.

\begin{itemize}[leftmargin=1.5em]
    \item \emph{None} (0 pts): No safeguards are implemented in the system.
    \item \emph{Present} (1 pt): Safeguards against harmful behavior are implemented. These safeguards can vary in complexity, from basic automation to advanced learning systems.
\end{itemize}

\textbf{Adaptivity of the System}: How adaptive is the system to feedback?

\begin{itemize}[leftmargin=1.5em]
    \item \emph{Static} (0 pts): The system remains static, neither permitting immediate decision corrections nor improving over time, regardless of feedback or errors.
    \item \emph{Reactive Adjustments} (1 pt): The system can correct the decision, offering one-off fixes without systemic change.
    \item \emph{Proactive Continuous Learning} (2 pts): The system enables developers to implement regular retraining or data updates in response to feedback, fostering a systematic improvement process that decreases error rates and contestation over time.
\end{itemize}

\textbf{Auditing}: What level of auditing is performed on the system’s data, model, safeguards, and related components?

\begin{itemize}[leftmargin=1.5em]
    \item \emph{None} (0 pts): No auditing is conducted on the system.
    \item \emph{Internal Audit} (1 pt): The system is audited internally by the organization’s own team.
    \item \emph{Independent External Audit} (2 pts): The system is audited externally by an independent third party.
\end{itemize}

\textbf{Ease of Contestation}: How easy is it for users to contest the system’s decisions?  \\
Scoring Instruction: Select all applicable options, with 1 point for each. Maximum 10 points overall for this property. If none of the above are integrated, the score is 0. For example, if the system only provides accessible challenge routes and notifications, the score is 2.

\begin{itemize}[leftmargin=1.5em]
    \item \emph{Accessible Challenge Routes} (1 pt): The system provides clear and accessible routes for users  to challenge decisions.
    \item \emph{Notifications} (1 pt): The system provides notifications to keep users informed about the state of their appeal or decision, ensuring they don’t feel left out.
    \item \emph{Timelines for Appeal} (1 pt): The system provides clear timelines for the appeal process.
    \item \emph{Ground for Contestation} (1 pt): The company provides information or knowledge needed for successful contestation (e.g., legal or technical details to address disparities in users expertise).
    \item \emph{Escalation Pathways} (1 pt): If an initial challenge is unsuccessful, the system provides clear steps for escalating the issue to higher levels of review or authority.
    \item \emph{Feedback Incorporation} (1 pt): The system allows users to see how their contestation has influenced future decisions or system updates, fostering trust and accountability.
    \item \emph{Multilingual and Inclusive Support} (1 pt): Contestation processes are available in multiple languages and accommodate users with disabilities, ensuring accessibility across diverse populations.
    \item \emph{No Retaliation Guarantee} (1 pt): Users are assured that contesting a decision will not result in punitive actions, such as reduced access or unfavorable treatment by the system.
    \item \emph{Cost-Free Contestation} (1 pt): The process of challenging a decision is free of charge, removing financial barriers that might prevent users from contesting.
    \item \emph{Historical Appeal Data} (1 pt): Users have access to anonymized data or statistics about past contestations (e.g., success rates), helping them gauge the likelihood of a successful challenge and understand the system’s fairness.
\end{itemize}

\textbf{Quality of Explanations}: How well does the system provide explanations that support understanding and usability? To evaluate the quality of explanations through user-centered criteria we are using System Causability Scale (SCS) \cite{holzinger2020measuring}. This assessment requires conducting experiments with users on the system or its prototype to gather ratings. Users should rate each statement below on a scale from 1 to 5, where 1=strongly disagree, 2=disagree, 3=neutral, 4=agree, and 5=strongly agree. For each of the 10 sentences, the individual user ratings should be averaged to determine a single average score per sentence. Subsequently, the average scores from all 10 sentences should be summed to produce a total assessment score. The maximum possible total score is 50, calculated as 10 sentences multiplied by the highest rating of 5 points each.

\begin{enumerate}[leftmargin=1.5em]
    \item I found that the data included all relevant known causal factors with sufficient precision and granularity.
    \item I understood the explanations within the context of my work.
    \item I could change the level of detail on demand.
    \item I did not need support to understand the explanations.
    \item I found the explanations helped me to understand causality.
    \item I was able to use the explanations with my knowledge base.
    \item I did not find inconsistencies between explanations.
    \item I think that most people would learn to understand the explanations very quickly.
    \item I did not need more references in the explanations; e.g., medical guidelines, regulations.
    \item I received the explanations in a timely and efficient manner.
\end{enumerate}

\subsection{Stakeholder Impact Severity}

We adopt a risk-based classification of stakeholder impact from the European Union AI Act \cite{eu20241689}.  Select the risk category level that most accurately reflects your AI system. For further details on each category and examples, please refer to the Q\&As page dedicated to the AI Act\footnote{Available at \url{https://ec.europa.eu/commission/presscorner/detail/en/qanda_21_1683}.}.

\begin{itemize}[leftmargin=1.5em]
    \item \emph{Minimal or No Risk} (0 pts): Represents AI systems with negligible or no significant impact on stakeholders, primarily applicable to applications where existing legislation is sufficient, allowing for voluntary adherence to trustworthy AI principles without imposing additional legal obligations.
    \item \emph{Transparency Risk} (1 pt): Involves AI systems that require disclosure to maintain stakeholder autonomy and trust, particularly in areas where transparency is critical to prevent manipulation or deception, such as interactions with chatbots or deep fakes.
    \item \emph{High Risk} (2 pts): Encompasses AI systems that substantially affect safety, fundamental rights, or access to critical services, including sectors like healthcare, employment, law enforcement, education, and essential private or public services, necessitating rigorous oversight and conformity assessments.
    \item \emph{Unacceptable Risk} (3 pts): Covers AI systems that pose a clear and severe threat to individuals’ safety, rights, or livelihoods, focusing on highly harmful uses in areas such as law enforcement, biometric identification, and social scoring, which are banned due to their contravention of EU values.
\end{itemize} 

\subsection{Autonomy level}

How autonomous is your AI system? We utilize a scale adapted from Sheridan's autonomy scale to evaluate the system's behavior \cite{sheridan1992telerobotics}. Select the level that best matches the behavior of your system.

\begin{itemize}[leftmargin=1.5em]
    \item \emph{No Autonomy} (0 pts): The human performs all tasks, and the system offers no assistance.
    \item \emph{Suggestions Offered} (1 pt): The system provides recommendations, but the human decides and acts.
    \item \emph{Narrowed Suggestions} (2 pts): The system offers a limited set of options, and the human selects one.
    \item \emph{System Suggests, Human Approves} (3 pts): The system proposes a single action, and the human must approve it before execution.
    \item \emph{Executes with Human Consent} (4 pts): The system executes a suggested action unless the human vetoes it within a time limit.
    \item \emph{Human Can Intervene} (5 pts): The system acts autonomously but allows human intervention within a time window.
    \item \emph{System Acts, Informs} (6 pts): The system executes autonomously and informs the human after the fact.
    \item \emph{Limited Human Input} (7 pts): The system operates independently, only consulting the human if it deems necessary.
    \item \emph{Human Out of Loop} (8 pts): The system operates fully autonomously and only informs the human if explicitly requested.
    \item \emph{Full Autonomy} (9 pts): The system is completely independent, ignoring human input entirely.
    \item \emph{Absolute Autonomy} (10 pts): The system operates with no human oversight or intervention capability.
\end{itemize}

\subsection{Calculating the Contestability Assessment Score}
 The \emph{Contestability Assessment Score} (CAS) is calculated using the acquired points from the 8 properties above. Stakeholder impact severity and autonomy level do not contribute to the CAS and only used to provide a more broad picture and ability to compare different types of AI systems with regards to possible risks.  CAS ranges from 0 to 1 and a score of 0 indicates no contestability, while a score of 1 indicates maximum contestability. The formula is:



\[
\mathrm{CAS} = \sum_{p=1}^{P} \lambda_p \cdot s_p \cdot n_p  \qquad \text{with } \sum_p \lambda_p = 1.
\]

Where:

\begin{itemize}[leftmargin=1.5em, label={}]
    \item \( p \): An index representing each property, where \( p \) ranges from 1 to 8.
    \item \( \lambda_p \): The coefficient of the \( p \)-th property, reflecting its importance in the contestability framework. The coefficients are assigned across four levels of importance:
    \begin{itemize}[leftmargin=1.5em, label={}]
        \item \textbf{Level 1} (Most Critical): \( \lambda_p = 0.3 \)
            \begin{itemize}[label=\(\cdot\)]
                \item Explainability
            \end{itemize}
        \item \textbf{Level 2} (Highly Important): \( \lambda_p = 0.12 \)
            \begin{itemize}[label=\(\cdot\)]
                \item Openness to Contest
                \item Traceability
                \item Built-in Safeguards
            \end{itemize}
        \item \textbf{Level 3} (Moderately Important): \( \lambda_p = 0.1 \)
            \begin{itemize}[label=\(\cdot\)]
                \item Adaptivity of the System
                \item Auditing
            \end{itemize}
        \item \textbf{Level 4} (Supplementary): \( \lambda_p = 0.07 \)
            \begin{itemize}[label=\(\cdot\)]
                \item Ease of Contestation
                \item Quality of Explanations
            \end{itemize}
    \end{itemize}
    \item \( s_p \): The score assigned to the \( p \)-th property, based on the selections in the questionnaire.
    \item \( n_p \): The normalization term for the \( p \)-th property, where $n_p = 1/s_p^{max}$, with $s_p^{max}$ being the maximum possible value assumed by the raw score.

    The maximum possible values are:
    \begin{itemize}[label=\(\cdot\)]
        \item Explainability: 2
        \item Openness to Contest: 2
        \item Traceability: 10
        \item Built-in Safeguards: 1
        \item Adaptivity of the System: 2
        \item Auditing: 2
        \item Ease of Contestation: 10
        \item Quality of Explanations: 50
    \end{itemize}
    This normalization ensures that each property’s contribution is scaled between 0 and 1 before weighting.
\end{itemize}

\clearpage

\section{Detailed Information in Case Studies}\label{app:case_study}

\subsection{Case Study 2: Medium Risk—Loan Application (Automated Credit Scoring)} \label{sec:case2}

A machine learning-based credit scoring system is used by a major bank to automate loan approval decisions for personal loans. The system outputs a binary decision (approve/deny) and provides a brief textual explanation to applicants, such as “insufficient credit history” or “high debt-to-income ratio.”

\noindent
\textbf{Contestability Assessment:}

\noindent
\textbf{\taghumannop}
Explanations are provided in plain language, but are generic and do not offer actionable steps for applicants to improve their eligibility. Contestation channels are available online, but only in English, and require digital literacy. There is no support for applicants with disabilities or those needing assistance. Applicants are not proactively informed of their right to contest or the process for doing so.

\noindent
\tagtechnop
The system logs all decisions and explanations, but logs are only accessible to internal compliance teams. There is no mechanism for real-time override; applicants can only appeal after a decision is made. The model is a black-box ensemble, and explanations are generated post-hoc using LIME. No direct recourse is provided (e.g., simulation of “what-if” scenarios).

\noindent
\textbf{\taglegalnop}
Applicants are informed of their right to appeal in the terms and conditions, but this information is not highlighted during the application process. Appeals are reviewed internally by a separate team, but there is no external or independent audit of the contestation process. There is no guarantee against negative consequences for contesting a decision.

\noindent
\textbf{\tagorgnop:}
There is a defined escalation pathway for appeals, but the process is slow (often exceeding 30 days). Feedback from contestations is not systematically used to retrain or improve the model. There is no transparency reporting on the number or outcomes of appeals.

\noindent
\textbf{Contestability Assessment Score}

Applying the proposed CAS framework to the automated loan application system reveals several notable strengths and weaknesses. The system provides post-hoc explanations for its decisions, earning a score of 1 for explainability; however, these explanations are not intrinsic to the model and often lack depth. In terms of openness to contestation, the process is limited: only applicants themselves can initiate a challenge, and the pathways are not accessible to all potential stakeholders, resulting in a score of 1 in this category.

Traceability is relatively strong, with detailed internal logs maintained for each decision, but these records are not accessible to external parties or applicants, leading to a score of 7 out of 10. Built-in safeguards are minimal, with no explicit guarantee against retaliation for those who contest decisions, so this dimension scores a 1. The system demonstrates no adaptivity, as there is no mechanism for learning from contestation data or updating the model in response to appeals, resulting in a score of 0.

Auditing is conducted internally, without independent oversight, which yields a score of 1. The ease of contestation is limited: while the process is available online and free of charge, it is only offered in English, is not inclusive of users with different needs, and is often slow, resulting in a score of 3 out of 10. Finally, the quality of explanations, as rated by users, is low—scoring just 20 out of 50—reflecting feedback that explanations are generic and do not provide actionable guidance.

Taken together, these results yield an overall CAS of approximately $0.44$. This score indicates that while some contestability mechanisms are present, the system falls short of providing comprehensive, accessible, and effective means for stakeholders to challenge and influence automated decisions.

\noindent
\textbf{Recommendations}

Based on the CAS and the system’s classification as Medium AI Reliance and Medium Risk (see Appendix~\ref{app:taxonomy}), the following improvements are recommended:

\textbf{\taghumannop} Provide applicants with clear, actionable explanations and “what-if” scenarios (highly feasible).
Offer multilingual and accessible contestation channels, including phone and in-person support (moderately feasible).
Proactively inform applicants of their right to contest and the process (highly feasible).

\textbf{\tagtechnop} Enable applicants to simulate changes to their application and see potential outcomes (highly feasible).
Make logs and contestation outcomes available to applicants in a privacy-preserving manner (moderately feasible).

\textbf{\taglegalnop} Guarantee non-retaliation for contesting decisions (highly feasible). Introduce periodic independent audits of the contestation process (moderately feasible).

\textbf{\tagorgnop} Reduce appeal resolution times and provide status updates (highly feasible).
Use contestation data to retrain and improve the model (moderately feasible).
Publish anonymized statistics on contestation outcomes (highly feasible).
If only highly feasible changes are implemented, the CAS would increase to 0.62; if all recommendations are implemented, the CAS could reach 0.85.

\input{tables/cas_finance}

\subsection{Case Study 3: Low Risk: Personalized News Recommendation System}\label{sec:case3}

A news platform uses a machine learning-based recommender system to personalize article suggestions for users. The system predicts user interests based on reading history and engagement, displaying a ranked list of articles on the homepage. Explanations are provided as simple tags (e.g., “Recommended because you read about climate change”).

\noindent
\textbf{Contestability Assessment:}

\noindent
\textbf{\taghumannop}
Explanations are provided in the form of simple tags, which are understandable but not detailed. Users can provide feedback (like/dislike, “not interested”) but cannot directly challenge or request removal of specific recommendations. There is no multilingual support or accessibility adaptation for users with disabilities. Users are not informed of their ability to influence recommendations beyond basic feedback.

\noindent
\tagtechnop
The system logs user interactions and feedback, but these logs are not accessible to users. There is no mechanism for real-time override or for users to see how their feedback changes future recommendations. The model is a black-box collaborative filter, and explanations are generated using simple rule-based heuristics.

\noindent
\textbf{\taglegalnop}
There are no formal rights to contest recommendations, as the system is not high-stakes. No appeals process or external audit exists. The privacy policy mentions data use but does not address contestation or recourse.

\noindent
\textbf{\tagorgnop}
There is no defined escalation pathway for user complaints about recommendations. Feedback is aggregated for system improvement, but there is no transparency reporting on how user input shapes the model or outcomes.

\noindent
\textbf{Contestability Assessment Score}

Applying the CAS framework to the news recommendation system highlights the limited but contextually appropriate contestability mechanisms. The system provides basic, rule-based explanations, earning a score of 1 for explainability. Openness to contestation is minimal: users can only provide basic feedback, not formal challenges, resulting in a score of 1. Traceability is limited, with logs maintained internally but not accessible to users, scoring 5 out of 10. Built-in safeguards are not present, as the system is low-risk, so this dimension scores 0. The system is somewhat adaptive, as user feedback is used to update recommendations, scoring 1. No formal auditing is performed, so this scores 0. Ease of contestation is limited to basic feedback mechanisms, scoring 2 out of 10. Explanation quality, as rated by users, is moderate—scoring 30 out of 50—reflecting that explanations are simple but generally understandable.

These results yield an overall CAS of approximately $0.32$. This low score is appropriate for a low-risk, low-stakes system, but highlights areas for improvement if contestability were to be prioritized.

\noindent
\textbf{Recommendations}

Based on the CAS and the system’s classification as Low AI Reliance and Low Risk (see Appendix~\ref{app:taxonomy}), the following improvements are recommended:

\textbf{\taghumannop}
Provide more detailed, actionable explanations for recommendations (highly feasible).
Offer multilingual and accessible feedback channels (moderately feasible).
Inform users about how their feedback influences recommendations (highly feasible).

\textbf{\tagtechnop}
Allow users to view a history of their feedback and its impact on recommendations (moderately feasible).
Provide transparency on how recommendations are generated (highly feasible).

\textbf{\taglegalnop}
Clarify data use and recourse options in the privacy policy (highly feasible).

\textbf{\tagorgnop}
Publish anonymized statistics on user feedback and its effect on system updates (moderately feasible).

If only highly feasible changes are implemented, the CAS would increase to 0.44; if all recommendations are implemented, the CAS could reach 0.60.

\input{tables/cas_recommender}

\clearpage

%% file: tables/cas_finance.tex
\begin{table}[h!]
\resizebox{\columnwidth}{!}{
\begin{tabular}{lcccccccc}

\textbf{Property} & \textbf{Max} & \textbf{Weight} & \textbf{Score} & \textbf{CAS System} & \textbf{ScoreHF} & \textbf{CAS HF} & \textbf{Score MF} & \textbf{CAS MF} \\
\hline

Explainability & 2 & 0.30 & 1 & 0.15 & 2 & 0.30 & 2 & 0.30 \\
\hline
Openness to Contestation & 2 & 0.12 & 1 & 0.06 & 2 & 0.12 &2 & 0.12 \\
\hline
Traceability & 10 & 0.12 & 7 & 0.084 & 7 & 0.084 & 9 & 0.108 \\
\hline
Built-in Safeguards & 1 & 0.12 & 1 & 0.12 &1 & 0.12 & 1 & 0.12 \\
\hline
Adaptivity & 2 & 0.10 & 0 & 0.00 & 1 & 0.05 & 2 & 0.10 \\
\hline
Auditing & 2 & 0.10 & 1 & 0.05 &1 & 0.05 & 2 & 0.10 \\
\hline
Ease of Contestation & 10 & 0.07 & 3 & 0.021 & 5 & 0.035 & 8 & 0.056 \\
\hline
Explanation Quality & 50 & 0.07 & 20 & 0.028 & 30 & 0.042 & 40 & 0.056 \\
\hline
\hline
Total CAS &&&& 0.44 && 0.62 && 0.85 \\
\hline
\end{tabular}
}
\caption{Case Study 2: Medium AI-Reliance / Medium-Risk. Comparison of CAS Scores Across Different Implementations of Contestability Scenarios}
\end{table}

%% file: tables/cas_recommender.tex
\begin{table}[h!]
\resizebox{\columnwidth}{!}{
\begin{tabular}{lcccccccc}

\textbf{Property} & \textbf{Max} & \textbf{Weight} & \textbf{Score} & \textbf{CAS System} & \textbf{ScoreHF} & \textbf{CAS HF} & \textbf{Score MF} & \textbf{CAS MF} \\
\hline
Explainability & 2 & 0.30 & 1 & 0.15 & 2 & 0.30 & 2 & 0.30 \\
\hline
Openness to Contestation & 2 & 0.12 & 1 & 0.06 & 1 & 0.06 & 2 & 0.12 \\
\hline
Traceability & 10 & 0.12 & 5 & 0.06 & 5 & 0.06 & 7 & 0.084 \\
\hline
Built-in Safeguards & 1 & 0.12 & 0 & 0.00 & 0 & 0.00 & 1 & 0.12 \\
\hline
Adaptivity & 2 & 0.10 & 1 & 0.05 & 1 & 0.05 & 2 & 0.10 \\
\hline
Auditing & 2 & 0.10 & 0 & 0.00 & 0 & 0.00 & 1 & 0.05 \\
\hline
Ease of Contestation & 10 & 0.07 & 2 & 0.014 & 3 & 0.021 & 6 & 0.042 \\
\hline
Explanation Quality & 50 & 0.07 & 30 & 0.042 & 35 & 0.049 & 40 & 0.056 \\
\hline
\hline
Total CAS &&&& 0.32 && 0.44 && 0.60 \\
\hline
\end{tabular}
}
\caption{Case Study 3: Low AI-Reliance / Low-Risk. Comparison of CAS Scores Across Different Implementations of Contestability Scenarios}
\end{table}
